\documentclass[]{aa}  % LaTeX A&A  Monotype Times Fonts
\usepackage{psfig}

\newcommand{\bcma}{$\beta$~CMa}
\newcommand{\acma}{$\alpha$~CMa}
\newcommand{\ecma}{$\epsilon$~CMa}
\newcommand{\ioni}{\,{\sc i}}
\newcommand{\ionii}{\,{\sc ii}}
\newcommand{\ioniii}{\,{\sc iii}}
\newcommand{\ioniv}{\,{\sc iv}}

\newcommand{\htot}{H$_{\rm tot}$}

\newcommand{\cmd}{cm$^{-2}$}
\newcommand{\cmt}{cm$^{-3}$}
\newcommand{\kms}{km\,s$^{-1}$}
\newcommand{\dens}{$n_{\rm e}$}

\begin{document}

   \thesaurus{09         % A&A Section 8: ISM
              (09.01.1;  % abundances
               09.01.2;  % atoms
               09.02.1;  % bubbles
               09.03.1;  % clouds
               08.09.2;  % stars: \bcma
               13.21.3)  % UV : ISM
             }
   \title{Diffuse ionized gas toward $\beta$ Canis Majoris
\thanks{Based on observations with the 
NASA/ESA \it Hubble Space Telescope, \rm obtained at the Space Telescope
 Institute, which is operated by the Association of Universities for 
Research in Astronomy, Inc., under NASA contract NAS5-26555.}}

   \author{Olivier Dupin, C\'ecile Gry}

   \offprints{O. Dupin}

   \institute{Laboratoire d'Astronomie Spatiale,
              B.P.8,
              13376 Marseille cedex 12,
              France,
	      (odupin@astrsp-mrs.fr, cecile@astrsp-mrs.fr)}

   \date{Received ; accepted }

   \maketitle

   \begin{abstract}
This paper presents the study of the interstellar medium toward $\beta$ CMa, a 
disk sight-line known for its low neutral gas density. This study uses  
high and medium resolution HST-GHRS spectra including lines from the 
following species~:
H\ioni, D\ioni, N\ioni, O\ioni, 
S\ionii, S\ioniii, Si\ionii, Si\ioniii, Si\ioniv, Al\ionii, Al\ioniii, 
Fe\ionii, Mg\ioni, Mg\ionii, Mn\ionii, C\ionii\ and C\ioniv.\\
The line of sight to $\beta$ CMa (153 pc) is dominated by two 
ionized regions with a velocity difference of 10~\kms. The ionized regions
account for most of the total hydrogen column density, around 2\,10$^{19}$~\cmd,
and the neutral gas represents only 10\% of the total gas.
The two ionized clouds display characteristics of the warm diffuse 
gas detected in the disk and the halo. 
Their gas-phase abundances indicate that their depletion is low, especially 
for the more ionized of the two clouds.\\
Special models of photoionization  by the two EUV-excess stars $\beta$ CMa and 
$\epsilon$ CMa would be needed for a detailed discussion of the 
ionizing mechanisms of the clouds ; their 
ionization ratios are nevertheless roughly compatible with 
collisional ionization at temperatures around 20\,000~K, substantially 
higher than the 
kinetic temperatures derived from the line widths. 
Their characteristics suggest that the clouds may 
be in the process of cooling down and recombining after having been
shocked and ionized by some violent events, possibly related to the
Local Bubble formation. 

   \end{abstract}

\section{Introduction}
Observations of the interstellar medium toward nearby stars show 
that the Sun is located in an interstellar void, often called the Local 
Bubble, filled with a hot ($10^6$~K) and low density (5 $10^{-3}$~\cmt) 
gas, responsible for part of the soft X-ray backround (e.g. Cox \& Reynolds 
1987 and references therein). The 
Bubble is irregularly shaped (Frish \& York 1983), extending to at least 50 
pc in most directions. {\it Copernicus} observations  (Gry et al. 1985)  
have shown an extension of the Bubble as far as the star \bcma\ 
($l=226^\circ$, $b=-14^\circ$) 153 pc from the Sun (Hipparcos
catalog). They report a 
neutral hydrogen column density from 1.0 to 2.2~$10^{18}$~\cmd, and a 
strongly ionized sight-line. Welsh (1991) and Welsh et al. (1994), 
have observed the Na\,{\sc i} line toward many neighbouring stars and confirmed 
the existence of this low density tunnel, 300 pc long and 50 pc wide,
in the direction of $\beta$ CMa.

More recently, Gry et al. (1995) analysed the UV absorption lines of the 
interstellar medium in the direction of \ecma\ ($l=239.8^\circ$,
$b_{II}=-11.3^\circ$), 132 pc from the Sun. They have shown that 
%this sight-line intercepts principally local matter~: 
the absorption lines are principally due to local matter~:
the Local Interstellar Cloud (LIC) in which the Sun 
is embedded (see Lallement \& Bertin 1992) as well as
a cloud already detected toward the very nearby (2.7 pc) star $\alpha$ CMa 
(Lallement et al. 1994), and they conclude that 
beyond three parsecs, the 
mean gas density is less than 4.5\,$10^{-4}$~\cmt. This makes this line of 
sight the most devoid region in the nearby interstellar medium. They also 
detected high ionization species like C\ioniv, attributed to the thermal 
conductive interface with the hot gas, thereby observed for the first time 
around two nearby diffuse clouds.

Observations of $\epsilon$ and $\beta$ CMa have been performed 
in the extreme UV 
with the {\it EUVE} spacecraft.  Due to their location in the 
tunnel and their EUV excess, \ecma\ and \bcma\ 
are respectively the first and the second 
brightest EUV sources in the sky, and the principal photoionization 
sources in the solar neighbourhood~: Vallerga \& Welsh (1995) have found a 
local interstellar hydrogen photoionization rate for $\epsilon$ CMa seven
times greater than previous estimates calculated for all nearby stars 
(Bruhweiler \& Cheng 1988). Modelling the EUV
stellar spectra, Cassinelli et al. (1995) and Cassinelli et 
al. (1996) found interstellar neutral hydrogen column 
densities of 1.0~$10^{18}$~\cmd\ and 2.0$\pm$0.2~10$^{18}$~\cmd\  
respectively for $\epsilon$ CMa and $\beta$ CMa, confirming the Gry et al. 
(1985) result for \bcma.

We present a study of the line of sight toward the 
B1 II-III 
star $\beta$ CMa, performed with ultraviolet spectra of the star.
Our purpose is to give a 
description of the structure and physical conditions of the 
ionized gas detected with Copernicus (Gry et al. 1985), and thereby discuss the
nature of diffuse clouds embedded in the Local Bubble.

Section 2 presents the data and describes their processing. In Sect. 3
a description of the line of sight toward $\beta$ CMa is given based on high 
and medium resolution data, and column densities and $b$-values are derived 
for each studied element and for each velocity
component wherever possible. Section 4 deals with the physical 
conditions in the clouds. The ionization of the clouds is treated in 
Sect. 5 and the gas phase abundances in 
Sect. 6. In Sect. 7, the nature of the gas is discussed in  comparison with 
other observations. 
\begin{table}
\begin{center}
\caption[]{Observed spectral ranges and atomic lines.}
\begin{tabular}{clcc}
\noalign{\smallskip}
\hline
\noalign{\smallskip}
range (\AA) & element  & wavelength (\AA) & $f$-value\\
\noalign{\smallskip}
\hline
\noalign{\smallskip}
\multicolumn {4}{c}{Ech-B data}\\
1805.0 -- 1814.6 & Si\ionii & 1808.013 & $2.0\,10^{-3}$\\
1841.6 -- 1875.8 & Al\ioniii & 1854.716 & $5.6\,10^{-1}$\\
 & Al\ioniii & 1862.789 & $2.8\,10^{-1}$\\
2338.5 -- 2350.7 & Fe\ionii & 2344.214 & $1.1\,10^{-1}$\\
2371.8 -- 2383.5 & Fe\ionii & 2374.461 & $2.8\,10^{-2}$\\
 & Fe\ionii & 2382.765 & $3.0\,10^{-1}$\\
2573.9 -- 2586.9 & Mn\ionii & 2576.877 & $3.5\,10^{-1}$\\
2792.1 -- 2807.1 & Mg\ionii & 2796.352 & $6.1\,10^{-1}$\\
 & Mg\ionii & 2803.531 & $3.1\,10^{-1}$\\
2846.7 -- 2860.7 & Mg\ioni & 2852.964 & $1.8\,10^{0}$\\
\noalign{\smallskip}
\multicolumn {4}{c}{G160M data}\\
1185.7 -- 1221.9 & S\ioniii & 1190.208 & $2.2\,10^{-2}$\\
 & Si\ionii & 1190.416 & $2.5\,10^{-1}$\\
 & Si\ionii & 1193.290 & $5.0\,10^{-1}$\\
 & N\ioni & 1199.550 & $1.3\,10^{-1}$\\
 & N\ioni & 1200.223 & $8.8\,10^{-2}$\\
 & N\ioni & 1200.710 & $4.4\,10^{-2}$\\
 & Si\ioniii & 1206.500 & $1.7\,10^{0}$\\
 & D\ioni & 1215.339 & $4.2\,10^{-1}$\\
 & H\ioni & 1215.670 & $4.2\,10^{-1}$\\
1229.8 -- 1265.9 & S\ionii & 1250.584 & $5.5\,10^{-3}$\\
 & S\ionii & 1253.811 & $1.1\,10^{-2}$\\
 & S\ionii & 1259.519 & $1.6\,10^{-2}$\\
 & Si\ionii & 1260.422 & $1.0\,10^{0}$\\
1300.1 -- 1336.0 & O\ioni & 1302.168 & $4.9\,10^{-2}$\\
 & Si\ionii & 1304.370 & $1.5\,10^{-1}$\\
 & C\ionii & 1334.532 & $1.3\,10^{-1}$\\
1390.4 -- 1426.2 & Si\ioniv & 1393.755 & $5.1\,10^{-1}$\\
 & Si\ioniv & 1402.770 & $2.6\,10^{-1}$\\
1524.1 -- 1559.3 & Si\ionii & 1526.707 & $2.3\,10^{-1}$\\
 & C\ioniv & 1548.195 & $1.9\,10^{-1}$\\
 & C\ioniv & 1550.770 & $9.5\,10^{-2}$\\
1644.4 -- 1679.3 & Al\ionii & 1670.787 & $1.8\,10^{0}$\\
\noalign{\smallskip}
\hline
\end{tabular}
\end{center}
\end{table}

\section{Observations and data reduction}
The observations described here have been obtained 
in December 1992 with 
the Goddard High Resolution
Spectrometer (GHRS) on board the Hubble Space Telescope (HST)
by Vidal-Madjar and 
collaborators. All lines above 1800 \AA \ have been observed with Ech-B,
and have a resolution of 85\,000 giving a velocity 
resolution of about 3.5~\kms. 
Because of the failure of 
Ech-A at that time, the short wavelength observations were made with the G160M, 
which gives a resolution of 20\,000 
and a velocity resolution of 15~\kms. Table 1 lists the observed 
spectral ranges and the spectral lines analysed.

All data were taken with the 0$\farcs$25 small science aperture 
(SSA), the procedure FP-SPLIT = 4 and a substepping of 2 samples per diode 
width (for details of the instrumentation see Duncan 1992). 
 For data processing, standard STSDAS procedures of the 
IRAF software were used. Wavelengths were assigned from the standard 
calibration tables. An error of $\pm$1 resolution element on the wavelength 
assignment is expected because of magnetic drift. 
We made use of an absorption line profile fitting software developed by 
Vidal-Madjar et al. (1977) also described in Ferlet et al. (1980a,b), 
and we checked our technique by comparing the results obtained with similar 
software developed by Welty (see Welty et al. 1991). Stellar lines were 
fitted with a low-order polynomial on both sides of each line 
and each interstellar absorption component was represented by the convolution 
of a theoretical Voigt profile with the instrumental profile. An iterative 
procedure which minimizes the quadratic differences between the profile and 
the data points allows to determine the column density of the 
absorbing elements {\it N} (cm$^{-2}$), the radial velocity of the cloud (\kms) 
and the line $b$-value (\kms) of each interstellar absorption component.
The instrumental profile is taken from Duncan (1992) and the atomic parameters 
from Morton (1991) except for the Si\ionii\,1808 line oscillator
strength which comes from Bergeson \& Lawler (1993).
\section{Analysis}
\subsection{The structure of the sight-line~: Ech-B data}
\begin{figure*}
\psfig{file=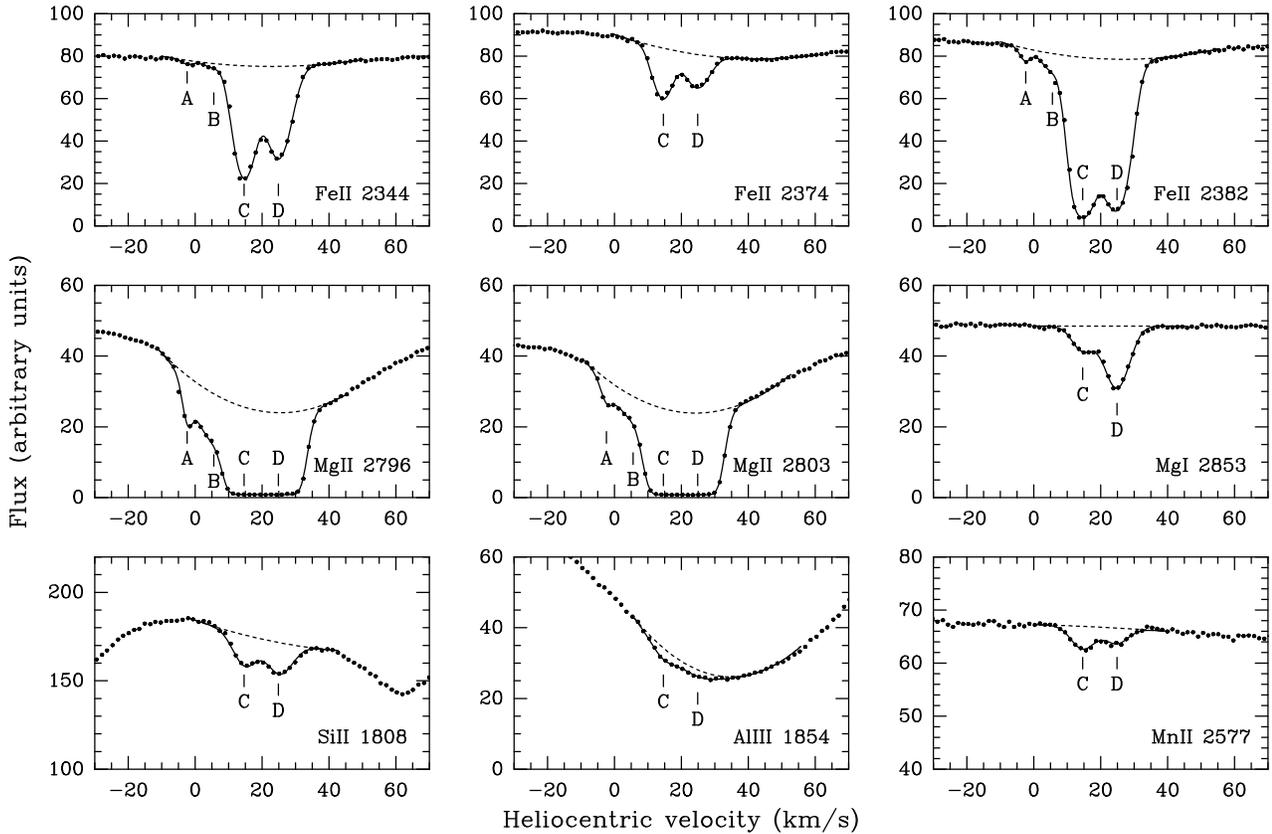,height=12.7cm,width=18cm,angle=-90}
\caption{Ech-B spectra of \bcma\ ($R\sim85\,000$). The points are the 
observations, the dashed lines the assumed stellar continua and the solid 
lines the fits to the interstellar absorption profiles. Column densities and 
$b$-values used for the fits are given in Table 2.}
\end{figure*}
The high resolution ($\sim$3.5~\kms) of the Ech-B data allows us to 
determine the velocity structure of the sight-line.
Two main components are clearly visible in all studied lines (Fig. 1). 
Two smaller features are detected only in the four strongest lines 
Fe\ionii\,2344, Fe\ionii\,2382, Mg\ionii\,2796 and Mg\ionii\,2803 (Fig. 1). 
These components are labelled A, B, C and D in order of increasing velocity.

In the analysis, the velocity shifts between components are constrained to be 
the same for all lines but the absolute velocities 
are allowed to vary slightly from one line to another to account for the 
uncertainty mentioned in Sect. 2. The derived velocities all lie within a
range of $\pm1.5$~\kms. The average of these velocities is therefore 
adopted, giving  $-2.5$, 5.5, 14.5 and 25~\kms\ for A, B, C and D respectively, 
and the spectra are shifted accordingly.
The above velocities thus define the heliocentric velocities of the 
components with an absolute precision of 3~\kms.

The Sun is known to be embedded in the Local Interstellar Cloud which is 
moving relative to it at a velocity of 26~\kms\ towards the direction 
$l=186^\circ$, $b=-16^\circ$ (Lallement \& Bertin 1992). The 
projection of its velocity vector on the $\beta$ CMa direction gives 20.3~\kms.
 This falls between component C and component D~: the absorption 
line due to the LIC is hidden and its study is therefore impossible at the 
resolution of the present data. It is nevertheless taken into account 
in the fit with the column density and $b$-value found by Gry et al. (1995) 
toward $\epsilon$ CMa. Another small component is present in the lines of sight 
of both \acma\ ('blue component', Lallement et al. 1994) and \ecma\ 
('component 2', Gry et al. 1995). 
Because of the vicinity of the \acma\ and \bcma\ sight-lines, this
cloud is likely to be present in the line of sight to \bcma\  
and it is also taken into account in the fit
with the \ecma\ column densities and $b$-values.

The component column densities and $b$-values are determined for all observed 
elements with the 
absorption profile fitting program mentioned in Sect. 2.
All the lines of an ion are fitted simultaneously.
The results are shown in Table 2. The error bars given take into account the
uncertainties in the stellar continuum modeling, the velocity positions of
the components and the parameters of the hidden components (LIC and 
\ecma\ component 2) as well as the statistical error of the data.
\begin{table*}
\begin{center}
\caption[ ] {Column densities and $b$-values of the individual clouds 
derived from the Ech-B spectra and when possible from the G160M data.}
\begin{tabular}{ccccccccc}
\hline
\noalign{\smallskip}
Cloud & \multicolumn {2}{c}{A} & 
\multicolumn {2}{c}{B} & \multicolumn {2}{c}{C} & \multicolumn {2}{c}{D}\\
Vel.& \multicolumn {2}{c}{$-2.5$~\kms}  & \multicolumn {2}{c}{5.5~\kms}  & 
\multicolumn {2}{c}{14.5~\kms}  & \multicolumn {2}{c}{25~\kms} \\
\noalign{\smallskip}
\hline
\noalign{\smallskip}
& {\it N} (\cmd) &{\it b} (\kms) &{\it N} (\cmd) &{\it b} (\kms) &
{\it N} (\cmd) &{\it b} (\kms) &{\it N} (\cmd) &{\it b} (\kms) \\
\noalign{\smallskip}
Fe\ionii &  1.4$\pm$.4 10$^{11}$ & 1.7$\pm$.8& 3.0$\pm$.1 
 10$^{11}$ & 2.8$\pm$.7& 1.2$\pm$.1 10$^{13}$ & 3.6$\pm$.1& 
1.0$\pm$.1 10$^{13}$ & 4.4$\pm$.2\\
\noalign{\smallskip}
Mn\ionii &  -& - & - & - & 2.2$\pm$.2 10$^{11}$ & "& 
 1.8$\pm$.2 10$^{11}$ & "\\
\noalign{\smallskip}
Mg\ionii &  3.4$\pm$.2 10$^{11}$ & 1.8$\pm$.2 & 1.5$\pm$.1 10$^{12}$& 
6.5$\pm$.2  & 3.1$\pm$.6 10$^{13}$ & 3.8$\pm$.2
& 5.9$\pm$.8 10$^{13}$& 4.6$\pm$.2\\
\noalign{\smallskip}
Mg\ioni & $\leq$3 10$^{9}$ & - & $\leq$1 10$^{10}$ & - & 
8.0$\pm$.6 10$^{10}$ & "& 2.5$\pm$.3 10$^{11}$ & "\\
\noalign{\smallskip}
Si\ionii &  - & - & - & - & 5.8$\pm$.5 10$^{13}$ & 3.75$\pm$.25& 
 7.5$\pm$.3 10$^{13}$ & 4.5$\pm$.3\\
\noalign{\smallskip}
Al\ioniii &  - & - & - & - & $1.9\pm$.3\ 10$^{11}$ & "&
 $1.8\pm$.6 10$^{11}$ & "\\
\noalign{\smallskip}
Si\ioniii & - & - &- & - & $\le5\ 10^{12}$ & "& 
$(1.5-10) 10^{14}$ & " \\
\noalign{\smallskip}
S\ionii & - & - &- & - & 1.1$\pm$.4 10$^{14}$ & "&
2.3$\pm$.5 10$^{14}$& "\\
\noalign{\smallskip}
N\ioni &  - & - &- & - & 5.5$\pm$.8 10$^{13}$ & 4.6$\pm$.6 & 
2.6$\pm$.5 10$^{13}$ & 5.5$\pm$.7\\
\noalign{\smallskip}
\hline
\end{tabular}
\end{center}
\end{table*}
\subsubsection{The Fe\ionii\ and Mn\ionii\ lines}
The three lines of Fe\ionii\ have different oscillator
strengths and two of them are not saturated, thus  they allow a reliable 
determination
of the column density and $b$-value of each component. 
The interstellar features are superimposed on weak 
stellar lines which can be easily modeled by polynomials. The Mn\ionii\,2577 
line is very weak, but the C and D components are clearly present. 
It lies in an 
area of the spectrum free of any stellar lines and the continuum is therefore
also very easy to determine. Because of the similarity of iron and manganese 
masses and with the assumption that Fe\ionii\ and Mn\ionii\ come 
from regions of the same 
temperature and turbulent velocity, we assume 
{\it b}(Fe\ionii)={\it b}(Mn\ionii) for C and D and fit these four lines 
simultaneously.
\subsubsection{The Mg\ionii\ and Mg\ioni\ lines}
The Mg\ionii\ lines are strongly saturated. It is thus in principle 
difficult to derive
column densities and $b$-values from the line profile fitting. 
However, with the assumption that {\it b}(Mg\ionii)={\it b}(Mg\ioni),
which is true if Mg\ioni\ and Mg\ionii\ arise from the same regions,
the two Mg\ionii\ lines 
can be fitted simultaneously with the unsaturated Mg\ioni\ line. 
Since the structure of the sight-line is 
known from the analysis of the Fe\ionii\ lines, the positions of all components
in the Mg\ionii\ profiles  can be determined precisely from the 
position of component A which is clearly visible in the spectra. 
Mg\ioni\ and Mg\ionii\ column densities and $b$-values could thus be derived 
for all components from the simultaneous fit of all the lines of the 
two species.

\subsubsection{The Si\ionii\ and Al\ioniii\ lines}
The Si\ionii\,1808 line is weak but the C and D components are 
clearly identified.
It is located in a portion of the spectrum with many narrow stellar lines. It 
is thus difficult to determine accurately the stellar continuum.
But the silicon $b$-value must be
between that of iron and that of magnesium, i.e.
{\it b}(Si\ionii)=3.5 -- 4.0~\kms\ for cloud C and 
{\it b}(Si\ionii)=4.2 -- 4.8~\kms\ for cloud D. With these constraints 
on the $b$-values only the deepest 
stellar profiles are permitted by the fit and the column density uncertainties
are thereby considerably  reduced.

In both
Al\ioniii\ lines, which appear in the same spectrum at 1854.7 and 1862.8 \AA,
two small features separated by $\simeq10.5$~\kms\ 
are detected at the bottom of the strong stellar lines. 
Since the atomic mass of aluminium is close to that of silicon, we adopt the 
Si\ionii\ $b$-values to fit the Al\ioniii\ lines.
\subsection{The G160M data}
\begin{figure*}[t]
\psfig{file=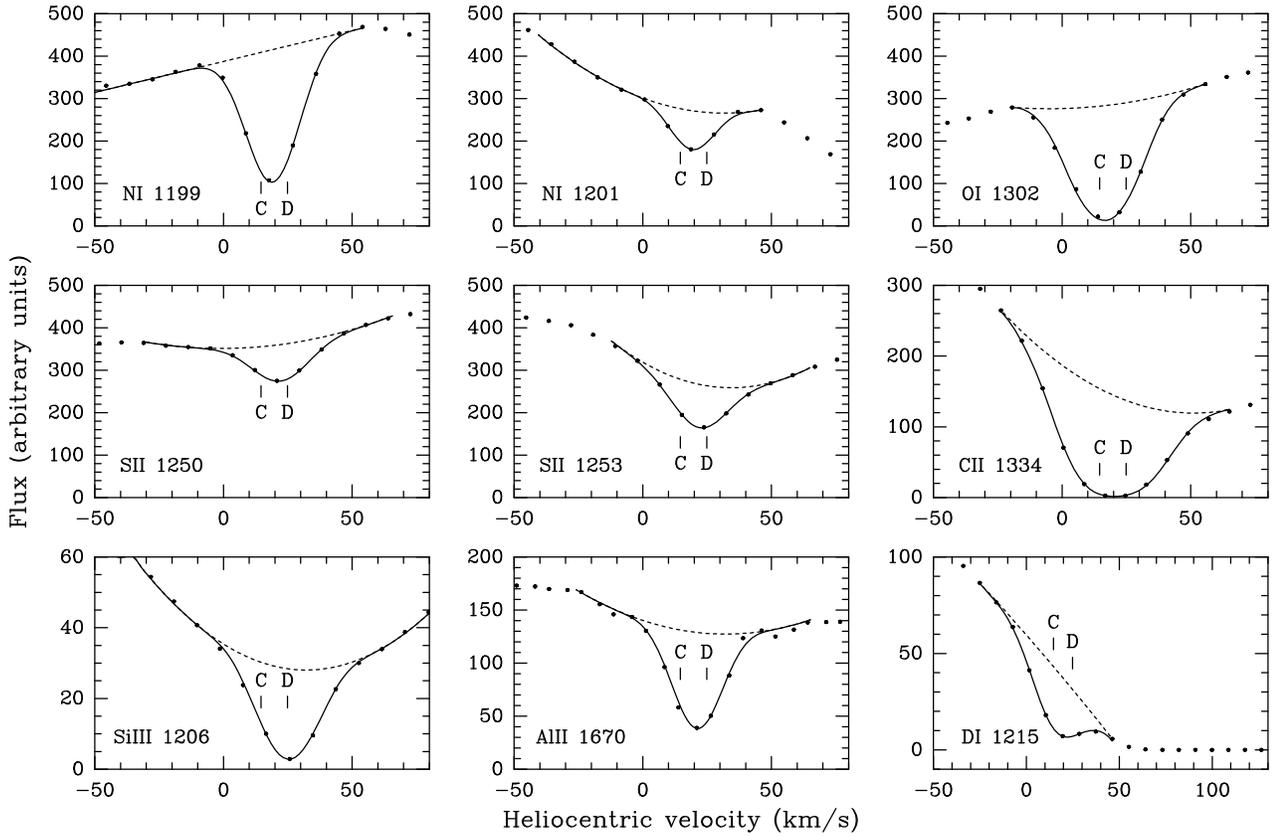,height=12.7cm,width=18cm,angle=-90}
\caption{Same as Fig. 1 for the G160M spectra of \bcma\ ($R\sim20\,000$). The
column densities used for the fits are given in Table 3. Although the position 
of components C and D are indicated in all spectra, the fits were 
performed with a
one-component model except for N\ioni, 
S\ionii\ and Si\ioniii.} 
\end{figure*}
Many important species only show absorption lines below 1800 
\AA. This is the case for neutrals like N\ioni, O\ioni\ and H\ioni\ as well as 
medium or highly ionized species like S\ionii, C\ionii, Si\ioniii, 
Si\ioniv\ and C\ioniv. Unfortunately, the Ech-A was not available at the time 
of the observations and the data were taken using the G160M grating with a 
lower resolution than the echelle gratings. With a resolution element of about
15~\kms, the different components seen in the Ech-B spectra are not resolved. 
An additional difficulty comes from the low rotational velocity of \bcma\ 
($v \sin i\sim30$~\kms) which implies relatively narrow stellar lines, 
sometimes not easily differentiated from the 
interstellar contribution. Nevertheless, for most 
species the total sight-line column density has been derived by fitting the 
interstellar absorption with one component and for N\ioni, S\ionii\ and
Si\ioniii, a multi-components fit was possible. Figure 2 shows some of 
the interstellar lines observed with G160M together with the best fits, and 
Table 3 summarizes the derived column densities.
\begin{table}
\begin{center}
\caption[ ] {Total sight-line column densities (cm$^{-2}$) derived from the 
G160M spectra.}
\begin{tabular}{ll}
\hline
\noalign{\smallskip}
N\ioni & 0.9 -- 1.1 10$^{14}$\\
O\ioni & $\geq5.5\ 10^{14}$\\
D\ioni & $\geq2.5\ 10^{13}$\\
Al\ionii & 5.0 -- 8.0 10$^{12}$\\
S\ionii & 3.1 -- 3.7 10$^{14}$\\
C\ionii & $\ge$4 10$^{14}$\\
Si\ioniii & 1.5 -- 10 10$^{14}$\\
Si\ioniv$^{1}$ & $\leq1.7\ 10^{11}$ ($\leq2.7\ 10^{11}$)\\ 
S\ioniii$^{2}$ & 0.9 -- 2.1 10$^{13}$\\
\noalign{\smallskip}
\hline
\noalign{\smallskip}
\multicolumn {2}{l}{$^{1}$ {\small 3$\sigma$ upper limit for $b=7$~\kms\ 
($b=11$~\kms)}}\\
\multicolumn {2}{l}{$^{2}$ {\small from Copernicus data, see Sect. 5.1}}\\
\end{tabular}
\end{center}
\end{table}
\subsubsection{Velocity calibration}
As mentioned in Sect. 2, the G160M spectra hold an uncertainty in the 
wavelength position of about one resolution element i.e. $\sim15$~\kms. 
However the velocity calibration has been improved with the help of the 
Si\ionii\ lines, since Si\ionii\ lines are present in one Ech-B spectrum 
(giving a precision of 3~\kms) and in 
four of the six G160M spectra (see Table 1). The convolution of the theoretical 
profile obtained for the high resolution Si\ionii\,1808 line with the
G160M line spread function gives a blended feature 
centered at a velocity of 20.5~\kms. This velocity is therefore adopted 
for the center of the interstellar Si\ionii\ absorption lines in the 
four G160M spectra, yielding a wavelength calibration of these spectra 
with the accuracy of the Ech-B spectra.
The velocity shifts thereby applied to the four G160M spectra rank from 
2.5 to 12.5~\kms, which are within the nominal precision of the data.
\subsubsection{The velocity-ionization relationship}
Figure 3 shows the central velocity of the absorption (from the one-component 
fit of the line) versus the ionization potential,
for all species for which a good calibration has been obtained via the above
process.
\begin{figure}
\psfig{file=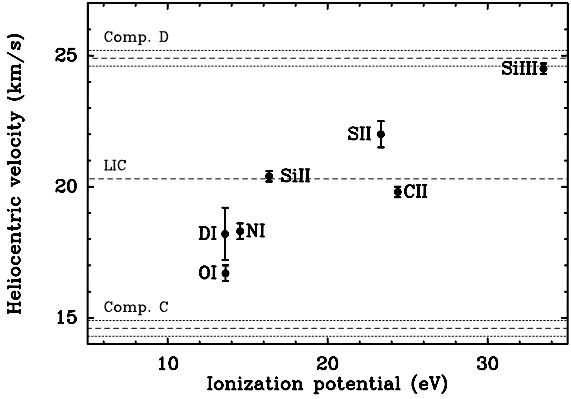,height=6cm,width=8.5cm,angle=-90}
\caption{Central velocity of the G160M interstellar features versus 
the ionization
potential of the elements. The dashed and dotted lines show the velocity 
positions of the clouds together with their relative uncertainty. 
The $\pm1.5$~\kms\ absolute error on the velocities is not shown 
since it affects all clouds simultaneously.}
\end{figure}
It shows that, as the ionization potential increases from D\ioni\ to Si\ioniii, 
the central velocity of the absorption profile increases from a velocity
intermediate to components C and D up to the velocity of component D, 
with a small departure 
for O\ioni\ and C\ionii\ which are saturated. This implies that component D 
contributes more than component C to the ionized species absorption and 
component C contributes 
more than component D to the neutrals absorption. 

In the following section 
we give some 
details on the analysis of each species. Special care has been put in 
trying to extract information on the distribution of matter between the 
two main components, although it is not always feasible at the G160M resolution.
\subsubsection{Neutral species~: N\ioni, O\ioni, H\ioni\ and D\ioni}
Three N\ioni\ lines are available in our data at 1199.5, 1200.2 and 
1200.7 \AA. The fact that the three lines have different oscillator 
strengths, with at least one of them  unsaturated, permits 
an accurate one-component fit, giving the following total 
line-of-sight N\ioni\ column density and $b$-value~: 
{\it N}(N\ioni)$=1.0\pm0.1\,10^{14}$~\cmd, {\it b}(N\ioni)$=8.5\pm0.3$~\kms. 

But a multi-component fit is also performed, taking into account four
components~: components C and D, which contribute both to the absorption 
feature according to Sect. 3.2.2, as well as the LIC and  component 2, the 
 column densities of which have been derived in the line of sight of \ecma\
with new Ech-A data (Gry \& Dupin, in prep.) and represent 25 \% of the 
N\ioni\ absorption in the \bcma\ sight-line.
The three N\ioni\ lines are fitted simultaneously with four components 
and the following constraints~: the velocity of each component is fixed at 
the known position, the $b$-values and column densities for the LIC and 
component 2 are those found for \ecma\, and the $b$-values for components C 
and D are between that of Mg and that given by the maximum 
temperature allowed for the clouds (see Sect. 4.1).

We derive 
{\it N}(N\ioni)$=5.5\pm0.8\,10^{13}$~\cmd, {\it b}(N\ioni)$=4.6\pm0.6$~\kms\ 
and 
{\it N}(N\ioni)$=2.6\pm0.5\,10^{13}$~\cmd, {\it b}(N\ioni)$=5.5\pm0.7$~\kms\ 
for component C and component D, respectively.

As the only available O\ioni\ line (1302 \AA) is saturated,
%it is inappropriate to fit it with a multi-component model. 
we could not find a unique solution with a multi-component model.
Therefore we 
simply derive a column density lower limit for the total sight-line using 
a one-component model, and taking the largest $b$-value that gives a 
reasonable fit. 
We find {\it N}(O\ioni)$\geq5.5\ 10^{14}$~\cmd\ for $b\leq11.0$~\kms.

The saturated Lyman $\alpha$ H\ioni\ interstellar
feature has large wings and it is difficult to separate it from the 
stellar line. Furthermore too many components have to be taken into account
to allow reliable results 
from profile fitting. But the Lyman $\alpha$
D\ioni\ line contribution can be separated from that of H\ioni\ and is 
only slightly saturated. We derive
{\it N}(D\ioni)$\geq2.5\ 10^{13}$~\cmd, $b\leq12.0$~\kms.
\subsubsection{The singly ionized species~: S\ionii, Si\ionii, C\ionii\ and 
Al\ionii}
Five lines of Si\ionii\ at 1190, 1193, 1260, 1304 and 1526 \AA\ are available
but the line at 1190.4 \AA\ is blended with the S\ioniii\, line at 1190.2 \AA\ 
and cannot be used. The 
two-component model derived from the analysis of the Ech-B Si\ionii\,1808 
line has been applied to the four available lines and gives a 
good fit for all of them.

The G160M data include three S\ionii\ lines at 1250, 1253 and 1259 \AA. 
The three lines have first been fitted simultaneously with one component only~:
this gives {\it N}(S\ionii)$=3.4\pm0.3\,10^{14}$~\cmd\ for the total sight-line.
As these transitions are weak and the lines are not saturated, 
most of the absorption must be due to the largest components.
The contribution of the LIC and component 2 can be estimated from 
their S\ionii\ column densities in the \ecma\ sight-line~:
Gry \& Dupin (1997) have found an upper limit of $6\ 10^{12}$~\cmd\ 
for the LIC, and  component 2 appears to be even 
weaker. These two clouds should therefore not 
account for more than 3\% of the sulfur in the \bcma\ sight-line.
The three S\ionii\ lines are thus expected to be 
strongly dominated by the two largest components C and D.
A two-component fit is thus performed, with the Si\ionii\ $b$-values, and gives 
{\it N}(S\ionii)$=1.1\pm0.4\,10^{14}$~\cmd\ for cloud C and 
{\it N}(S\ionii)$=2.3\pm0.5\,10^{14}$~\cmd\ for cloud D. 
It is seen that the sum of the two cloud contributions is equal to the 
one-component fit result, but with a larger uncertainty.

The C\ionii\,1334 line is strong and saturated. Each component of the 
sight-line gives a significant contribution to the interstellar absorption 
feature and, as for O\ioni,  
we only derive a column density lower limit~: N(C\ionii)$\geq4\,10^{14}$~\cmd\ 
with
$b\leq14.5$~\kms. The interstellar C\ionii *1335 line is located at the end of
the spectrum and it was not possible to use it.  

The Al\ionii\ line at 1670 \AA\ is located in a spectral region which does not 
include Si\ionii\ lines 
and thus does not have a precise velocity calibration. It is therefore 
fitted with a one-component model which gives 
{\it N}(Al\ionii)$=6.5\pm1.5\,10^{12}$~\cmd\ for 
{\it b}$=6.5\pm0.7$~\kms. The large uncertainty in the column density is due
to the fact that we have only one Al\ionii\ line and that the stellar 
continuum is not well defined at these wavelengths.
\subsubsection{The highly ionized species~: S\ioniii, Si\ioniii, Si\ioniv\ and 
C\ioniv}
The S\ioniii\ line at 1190.2 \AA\ is unfortunately severely blended with the 
Si\ionii\ line at 1190.4 \AA. Thus, no
satisfactory stellar continuum could be drawn and this line was not used.

The Si\ioniii\,1206 line is saturated but the stellar line is well defined and
can be modelled by a parabola. The fit confirms that most of the 
Si\ioniii\ is in cloud D and with the assumption that 
{\it b}(Si\ioniii)={\it b}(Si\ionii), gives 
{\it N}(Si\ioniii)$=1.5$ -- $10\,10^{14}$~\cmd\ for cloud D and an upper
limit for cloud C of {\it N}(Si\ioniii)$\leq5\,10^{12}$~\cmd. The large 
uncertainty on the D column density is due to the line saturation. 

A spectral region containing the Si\ioniv\ lines at 1393 \AA\
and 1402 \AA\ was observed, but no interstellar absorption
feature was detected at these wavelengths. 
Only 3$\sigma$ upper limits are derived for the 
column density. If it is assumed that Si\ioniv\ comes from the clouds
themselves i.e. {\it b}(Si\ioniv)={\it b}(Si\ionii)$\sim7$~\kms, 
the strongest line implies 
{\it N}(Si\ioniv)$\leq1.7\,10^{11}$~\cmd. If Si\ioniv\
is produced by a high temperature interface between the clouds and the hot
surrounding medium, a 3 $\sigma$ upper limit gives 
{\it N}(Si\ioniv)$\leq2.7\,10^{11}$~\cmd\ for $T=2\,10^{5}$~K.

\begin{figure}
\psfig{file=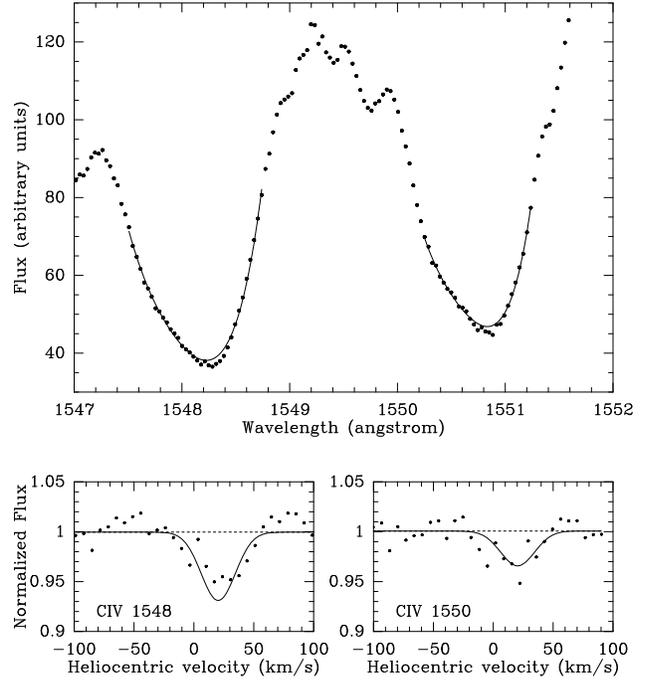,width=8.8cm,angle=0}
\caption{Possible interstellar absorption in the C\ioniv\ lines. {\it Upper}~: 
the \bcma\ spectrum between 1548 and 1552 \AA. The solid
lines are fourth order polynomials fitting the stellar lines. Possible
interstellar absorptions appear at the bottom of the lines. {\it
Lower}~: model of the LIC absorption found for $\epsilon$ CMa (Gry et al. 1995)
superimposed on the normalised \bcma\ spectra. 
%The points are the data, the dashed lines the
%stellar continua, and the solid lines the model.
}
\end{figure}
Fig. 4 shows the \bcma\ spectrum between 1547 \AA\ and 1552 \AA\ containing
the C\ioniv\ 1548,1550 doublet. When the stellar lines are represented
by fourth order polynomials, an absorption feature appears at the bottom of
both stellar lines. Although the interstellar nature of these features is 
not beyond dispute they nevertheless occur in both lines at the same velocity
which also happens to be the LIC velocity ($\sim20$~\kms).
In their study of the interstellar gas toward
\ecma, Gry et al. (1995), reported a C\ioniv\ absorption at the LIC velocity, 
with a width corresponding to thermal broadening at $T=2\ 10^5$~K. 
They suggest 
that the absorption could be due to the interface between the LIC and 
the hot surrounding medium. In Fig. 4,
the C\ioniv\ absorption lines in \bcma\ are compared to the model found for 
the LIC absorption in the case of \ecma\ 
({\it N}(C\ioniv)$=3.2\ 10^{12}$~\cmd, $T=2\ 10^{5}$~K, $V=20.3$~\kms). 
The C\ioniv\ data are consistent with the expected LIC absorption 
as determined toward \ecma.
\subsection{Comparison with Copernicus observations}
\bcma\ has been observed with Copernicus (Gry et al. 1985) at shorter 
wavelengths
and with a slightly higher resolution than that of G160M, but not
sufficient to resolve the different components. The Copernicus O\ioni\ and 
Si\ionii\ column densities are compatible with the G160M values. 
But the N\ioni, D\ioni, S\ionii\ and Si\ioniii\ column 
densities derived from the Copernicus observations are systematically  
lower than those we derive from our data. 
We have therefore reviewed the Copernicus background estimates used by 
Gry et al. and reached the conclusion that the contamination due to diffuse
light in the Copernicus instrument may have been incorrectly estimated.
This is clear in the case of the Si\ioniii\ line at
1206.5 \AA\ which 
is undoubtedly saturated in our data (the background level is given by the 
bottom of the
saturated Lyman $\alpha$ line present in the same spectrum) while it does not
appear saturated in the Copernicus spectrum. According to 
Rogerson et al. (1973), the background of the Copernicus data is roughly 
25\% of the continuum. If this background contribution has been underestimated,
the Copernicus column density of unsaturated lines may 
have been underestimated by 25\%.
If such a correcting factor is allowed, Copernicus and G160M values are in
agreement for N\ioni\ and S\ionii. Since the Si\ioniii\,1206 line 
is saturated, the column density estimate is very sensitive to any background 
misplacement and the value derived from Copernicus had been more strongly 
underestimated.
\section{Physical conditions in the clouds}
\subsection{Temperature and turbulent velocity}
Thanks to the mass difference between Mg and Fe,
the thermal and turbulent contributions to the
$b$-values can in principle be disentangled, providing 
a range of temperatures and turbulent velocities 
for each cloud. The $b$-values are listed in Table 2. For component B, 
the $b$-values found for Mg\ionii\ and Fe\ionii\
are so different that the 
ranges do not allow a measure of the turbulence.
Conversely, for components A, C and D, the ranges for Mg\ionii\ and Fe\ionii\
overlap and as the atomic mass of these elements imposes
{\it b}(Mg\,{\sc ii})$\geq${\it b}(Fe\,{\sc ii}), this only provides 
upper limits for the temperatures~: $T\leq 8000$~K for
cloud A, $T\leq9500$~K for cloud C and $T\leq 13\,500$~K for cloud D. 
The derived ranges for the turbulent velocities are the following~:
1.7~\kms\ $\leq V_{\rm turb}\leq3.3$~\kms\ for cloud A, 
3.2~\kms\ $\leq V_{\rm turb}\leq4.0$~\kms\ for cloud C 
and 4.0~\kms\ $\leq V_{\rm turb}\leq4.8$~\kms\ for cloud D. 
\subsection{Electron density}
Cloud C and cloud D are both detected in Mg\ioni\ and Mg\ionii. If 
ionization equilibrium between Mg\ioni\ and Mg\ionii\ following e.g.
Frisch et al. (1990) and Lallement et al. (1994) is assumed, the
electron density in the clouds can be derived~: 
\dens$ = \Gamma/(\alpha_{\rm rad} + \alpha_{\rm die}) *${\it N}(Mg\ionii)/{\it
N}(Mg\ioni)$ - (k_{\rm exch}+C)$~; where the Mg\ionii\ radiative
recombination rate 
($\alpha_{\rm rad}$) is taken from Aldrovandi \& P\'equignot (1973), the
dielectronic recombination rate ($\alpha_{\rm die}$) from Nussbaumer \& Storey
(1986), the Mg\ioni\ ionization rate by charge exchange 
with protons ($k_{\rm exch}$) from Allan et al. (1988), 
and the MgI radiative ionization rate ($\Gamma$) from Phillips et al. (1981).
{\it C} is the Mg\ioni\ 
collisional ionization rate but it is negligible at the 
temperatures involved.  For a warm cloud typical temperature of $T\sim7000$~K,
the range of {\it N}(Mg\ionii)/{\it N}(Mg\ioni) values derived from the 
observations implies \dens=0.18 -- 0.54~\cmt\ for cloud C and 
\dens=0.38 -- 1.82~\cmt\ for cloud D.
\section{Ionization of the clouds}
\subsection{The ionization degree}
\begin{table*}
\caption[]{Ionization ratios derived from the observations and from the
collisional ionization model of Sutherland \& Dopita (1993).}
\begin{center}
\begin{tabular}{lllllll}
\hline
\noalign{\smallskip}
 & \multicolumn{2}{c}{observed} & \multicolumn{4}{c}{Sutherland \& Dopita (1993)}\\
\noalign{\smallskip}
\hline
\noalign{\smallskip}
 Ratios & C & D & $T$=15\,000~K & $T$=20\,000~K & $T$=25\,000~K & 
$T$=30\,000~K\\
\noalign{\smallskip}
\hline 
\noalign{\smallskip}
H\ioni/H\ionii & $\leq1.5$ & $\leq0.07$ &
1.79 & 7.2 $10^{-2}$ & 1.2 $10^{-2}$ & 3.4 $10^{-3}$\\
\noalign{\smallskip}
Si\ionii/Si\ioniii & $\geq10.6$ & 0.07 -- 0.52 & 47 & 1.1 & 0.17 & 
8.5 $10^{-2}$\\
\noalign{\smallskip}
Si\ioniii/Si\ioniv & - & $\geq600$ & - & 3.8 $10^{4}$ &
648 & 60\\
\noalign{\smallskip}
Si\ionii/Si\ioniv & $\geq288$ & $\geq212$ & - & 4.1 $10^{4}$ & 134 & 5.46\\
\noalign{\smallskip}
S\ionii/S\ioniii & - & 8.6 -- 33 & 5300 & 30.4 & 3.1 & 1.0\\
\noalign{\smallskip}
Fe\ionii/Fe\ioniii & $\geq0.14$ & $\geq0.05$ & 19.6 & 0.52 & 0.13 & 
7 $10^{-2}$\\
\noalign{\smallskip}
Mg\ionii/Mg\ioniii & $\geq0.20$ & $\geq0.20$ & 7.7 & 0.24 & 
3.3 $10^{-2}$ & 9.3 $10^{-3}$\\
\noalign{\smallskip}
N\ioni/N\ionii & $\geq0.09$ & $\geq0.02$ & 18.8 & 0.18 & 
2.9 $10^{-2}$ & 1.2 $10^{-2}$\\
\noalign{\smallskip}
Al\ionii/Al\ioniii & $\leq50$ & $\leq67$ & 2.1 $10^{3}$ & 62.4 & 8.8 & 2.38\\
\noalign{\smallskip}
\hline
\end{tabular}
\end{center}
\end{table*}
Sulfur is a good element to use to derive the total hydrogen abundance
because it is only slightly (Harris \& Mas Hesse 1986) or not (Sofia et al. 
1994) depleted in the interstellar gas. We thus apply the solar abundance 
given by Anders \& Grevesse (1989), $\log[N({\rm S})/N({\rm H})]=-4.73$, to 
our determined sulfur column densities to derive the total hydrogen column 
density in the main components C and D. 
 The S\ioniii\ column density is taken from 
Gry et al. (1985), 
{\it N}(S\ioniii)=0.9 -- 2.1 10$^{13}$~\cmd, after correction for the likely 
background misplacement as explained in Sect.
3.3. Most of the S\ioniii\ is assumed to be in component D, as is Si\ioniii\ 
(see Sect. 3.2.5), thus for cloud D, the total column density is derived from 
the sum of S\ionii\ and S\ioniii\ column densities~: 
$N({\rm H}_{\rm tot})=1.0$ -- $1.6\ 10^{19}$~\cmd. For cloud C, the S\ionii\
column yields $N({\rm H}_{\rm tot})=3.7$ -- $8.0\ 10^{18}$~\cmd.

For the total sight-line, the sum of the total sight-line S\ionii\ 
column density and the above S\ioniii\ column density gives 
$N({\rm H}_{\rm tot})=1.7$ -- $2.1\ 10^{19}$~\cmd. 
Cassinelli et al. (1996) have derived {\it N}(He\ioni)$\geq1.4\,10^{18}$~\cmd\ 
from the He\,{\sc i}\,504 \AA\ line. With a
cosmic abundance of 0.1 for He this implies $N({\rm H}{_{\rm tot}}) 
\geq 1.4~10^{19}$~\cmd\ which is in agreement with our value.

The neutral hydrogen column density toward $\beta$ CMa has been estimated 
by Gry et al. (1985) by fitting the Lyman-$\beta$ profile observed with
Copernicus. They found {\it N}(H\,{\sc i})$=1.6\pm0.6\ 10^{18}$~\cmd. Recently, 
Cassinelli et al. (1996), using EUVE observations, have modelled the 
stellar spectrum and found an interstellar neutral hydrogen absorption of 
{\it N}(H\,{\sc i})$=2.0\pm0.2\,10^{18}$~\cmd, which is in good agreement with
the previous estimate. 

If we adopt the wider range for {\it N}(H\ioni), 
the neutral fraction is {\it N}(H\ioni)/{\it N}(\htot)=0.05 -- 0.13
for the total line of sight.  This implies that about 90\% of 
the matter of the sight-line is ionized.

The share of neutral hydrogen between the main components is
difficult to assess. As seen in Table 2, the way N\ioni\ 
is distributed over the 
main components is very different from the way Mg\ionii\ is distributed, 
although
the ionization potentials of the two elements are close (14.5 and 15.0~eV)~:
the column density ratio between C and D is close to 2 for N\ioni\ and close to 
1/2 for Mg\ionii. This suggests that the share of neutral hydrogen between
C and D cannot be derived from that of neutral nitrogen.
Nevertheless, since component D is in general the more ionized,
and since  the N\ioni\ ionization potential is slightly larger than that of
H\ioni, it is likely that the fraction of H\ioni\ contained in D 
does not exceed the fraction of N\ioni, which is 20\% to 30\%
of the total neutral nitrogen in the line of 
sight (Sect. 3.2.3). It is thus likely that no more than 30\% of the 
total neutral gas lies in
cloud D~: {\it N}(H\ioni)$\le6.6\ 10^{17}$~\cmd. 
This gives {\it N}(H\ioni)/{\it N}(\htot)$\leq7$\% for D. 

As for component C, what we can safely say is that it does contain a good 
fraction of the neutral gas since the D\ioni\ and O\ioni\ features are centered
at a velocity close to that of component C, and that it contains  no more
H\ioni\ than that present in the total line of sight. This implies 
that cloud C is partially ionized with 
{\it N}(H\ioni)/{\it N}(\htot)$\leq60$\%.
\subsection{Origin of the ionization}
Table 4 summarizes all the information available on ionization 
ratios in the
clouds or in the line of sight as a whole. The case of hydrogen has been
discussed above. For sulfur, since all S\ioniii\ is assumed to be in  
cloud D,
{\it N}(S\ionii)/{\it N}(S\ioniii) is derived for
cloud D only. For iron and magnesium, upper limits for their
total gas phase abundances {\it N}(X)$_g$ were derived by adopting the 
minimum depletion values found in the
interstellar medium as given by Fitzpatrick (1996, 1997). Then upper
limits on their second ion column density were derived 
by {\it N}(X\ioniii) = {\it N}(X)$_g$ 
- {\it N}(X\ionii). This yields lower limits for the ratios 
{\it N}(X\ionii)/{\it N}(X\ioniii). For nitrogen, we proceed similarly but
with the first and second ion column densities. For aluminium, upper limits on 
the {\it N}(Al\ionii)/{\it N}(Al\ioniii) ratio are derived taking the total
sight-line Al\ionii\ column density as an upper limit on the individual cloud
Al\ionii\ column densities.  For silicon, the ionization ratios are derived
directly from the data, taking the same upper limit for the 
Si\ioniv\ column densities of the individual clouds as that derived for the
total Si\ioniv\ column density.

In principle these ionization ratios allow a discussion of the 
ionization mechanisms involved in the gas. 

The most straightforward mechanism is
photoionization by the star \bcma\ as well as its neighbour \ecma\,
which is, as already mentioned, the strongest ionizing source in the
Solar neighbourhood (Vallerga \& Welsh 1995). But the
comparison of our ionization ratios with models and other observations
found in the litterature is not really conclusive.

In general, H\ioni\ regions are dominated by singly ionized species 
like Si\ionii, S\ionii\ and Al\ionii\ while photoionized H\ionii\ regions show 
strong absorptions in Si\ioniv, S\ioniii\ and Al\ioniii\ 
(e.g. Fitzpatrick \& Spitzer 1994, Spitzer \& Fitzpatrick 1995, Jenkins \& 
Wallerstein 1996, Savage \& Sembach 1996a,b). 
The \bcma\ line of
sight main components are thus atypical because, while hydrogen is 
substantially ionized, S\ionii, Al\ionii\ and Si\ionii\ (for cloud C) or
Si\ioniii\ (for cloud D) are still dominant and Si\ioniv\ is too weak to 
be detected.
Cowie et al. (1981), in a study of the origin and the distribution of
C\ioniv\ and Si\ioniv\ with 46 O-B stars observed with IUE, found that
almost all stars are associated with C\ioniv\ and Si\ioniv\ column densities of
$\sim10^{13}$~\cmd.
They conclude that most of this absorption arises in photoionized H\ionii\ 
regions. They also present a simple photoionization model predicting the 
H\ionii\ regions 
C\ioniv\ and Si\ioniv\ column densities. For a 25\,000~K effective temperature
star and an electron density of 1~\cmt\ it gives {\it N}(C\ioniv)$=2.6\
10^{10}$~\cmd\ and {\it N}(Si\ioniv)$=3.7\ 10^{12}$~\cmd\, the latter being
a factor of 10 higher than the \bcma\ upper limit.  The model gives values
compatible with the \bcma\ data for electron densities of 0.01~\cmt\ or less,
much smaller than the range of values we derive (see Sect. 4.2).

York (1983) found evidence for H\ionii\ regions in the line of sight to
$\lambda$ Sco (B2 IV),
for which he compared the ionization ratios with those given by
photoionization models with radiation fields of stars of different
effective temperatures. For the effective temperature of \bcma\
($\sim25\,000$~K, Cassinelli et al. 1996), he quotes
the results from Thuan (1975) for an electron density of 0.1~\cmt~:
S\ionii/S\ioniii$=8$ and Si\ionii/Si\ioniii$=1.6$.
In component C, Si\ionii\ is clearly less ionized than in the model.
For component D S\ionii/S\ioniii\ is compatible with the quoted model value
and Si\ionii\ is somewhat more ionized, which  could be explained in
view of the EUV excess of \bcma\ and \ecma. On the other hand,
Spitzer \& Fitzpatrick (1995) have identified a
photoionized H\ionii\ region toward HD\,149881, a halo star of similar 
spectral type (B0.5\,III), where S\ionii/S\ioniii\ is less than 1, i.e. 
similar to that of 'typical'
H\ionii\ regions around hotter stars (e.g. the three H\ionii\ regions detected
in the line of sight of the O9\,I+WC8 star
$\gamma^{2}$ Velorum by Fitzpatrick \& Spitzer 1994) and significantly lower
than the ratio we measure for component D.

Clearly the comparison of all these observations shows that a more detailed
ionization model by the stars is required before any conclusion can be drawn.
However, detailed photoionizing models by these two sources which  take into
account their EUV excess evidenced by EUVE (Cassinelli et al. 1995,
Cassinelli et al. 1996) is beyond the scope of this paper.

\begin{table*}
\caption[]{Gas phase abundances in the clouds and for the total sight-line}
\begin{center}
\begin{tabular}{lcccccccc}
\hline
\noalign{\smallskip}
 & [C/H] & [N/H] & [O/H] & [Mg/H] & [Si/H] & [Al/H] & [Mn/H] & [Fe/H]\\
\noalign{\smallskip}
\hline
\noalign{\smallskip}
cloud C & - & $\ge-0.5$ & - & $\ge-1.1$ & $-0.55^{+0.23}_{-0.18}$ & $\leq-0.13$
& $\ge-1.13$ & $\ge-1.37$\\
\noalign{\smallskip}
cloud D & - & $\ge-0.55$ & - & $\ge-1.1$ & $+0.15^{+0.33}_{-0.56}$ & 
$\leq-0.57$ & $\ge-1.53$ & $\ge-1.76$\\
\noalign{\smallskip}
total & $\geq-1.28$ & $-0.25^{+0.24}_{-0.15}$ & $\geq-0.52$ & $\geq-1.00$ &
$+0.02^{+0.26}_{-0.45}$ & $-0.92^{+0.14}_{-0.16}$ & $\ge-1.30$ & $\ge-1.49$\\
\noalign{\smallskip}
\hline
\end{tabular}
\end{center}
\end{table*}

Our ionization ratios can also be compared with
those given by
 collisional ionization. The comparison in Table 4 with the ratios given
for various
equilibrium temperatures in the model of Sutherland \& Dopita (1993) shows
that for component D there is an agreement  for all ratios for 
temperatures around $T\sim21\,000$~K.
For component C, although the temperature ranges derived from
Si\ionii/Si\ioniii\ and Al\ionii/Al\ioniii\ do not overlap, the
ratios point to temperatures ranging from about $16\,000$~K to about
$22\,000$~K.
For both components, these temperatures are significantly higher than the
kinetic temperatures of the
clouds derived from profile fitting (Sect. 4.1). Thus if the clouds are
collisionally ionized, they
cannot be in equilibrium, which indicates that they are
still in a cooling and recombining phase.

Trapero et al. (1996) in their study of halo high velocity clouds
(HVCs) with the GHRS also proposed that the ionization of several HVCs in the
line of sight to 23\,Ori and $\tau$\,CMa is due to collisional ionization.
As in the case of the clouds in the \bcma\ sight-line they are
not in ionization equilibrium and are overionized with respect to
their kinetic temperatures~: the ionization balance
corresponds to a collisional ionization equilibrium near $T=25\,000$~K while
the temperature derived from the Mg\ionii\ lines is $\le12\,000$~K.
Trapero et al (1996) suggest that
the HVCs consist of already cooling warm ionized gas, originally probably 
collisionally ionized and heated by a shock. 
Our results tend to support this interpretation for
components C and D,
in view of the depletion results which are discussed in Sects. 6 and 7.
\section{Gas phase abundances}
The gas phase
abundance of an element X relative to H is defined by 
$$\left[\frac{\rm X}{\rm H}\right] = 
\log\left[\frac{N({\rm X})}{N({\rm H})}\right] - 
\log\left[\frac{N({\rm X})}{N({\rm H})}\right]_{\rm cosmic}$$
where $\log\left[N({\rm X})/N({\rm H})\right]_{\rm cosmic}$ is the 
solar abundance as given by Anders \& Grevesse (1989). 

All available [X/H] derived from our 
study are listed in Table 5. When possible their values for 
components C and D are listed separately. All relevant 
ionization stages are considered and when applicable (Si and Al) 
we use the sum of 
the various stage column densities.
{\it N}(O\ioni) is compared to {\it N}(H\ioni), as is {\it N}(N\ioni)
for consistency with other studies although in the case of N\ioni, 
as mentioned in Sect. 5.1, some ionization effects might be involved.
All other element column densities are compared to total hydrogen 
column density.
As Si\ioniii\ is the dominant stage in component D, it is probable that 
Fe\ioniii, Mg\ioniii\ and Mn\ioniii\ are also significant. This is why 
in Table 5 the gas phase abundance of their singly ionized form is 
listed as a lower limit only.

It appears that in cloud D silicon is not at all, or only slightly, depleted. 
Its gas phase
abundance is at least 0.44 dex higher than the value given in the Table 5
of Jenkins \& Wallerstein (1996) for the depleted low velocity gas in the disk.
In their Fig. 6, Savage \& Sembach (1996a) have plotted the mean gas phase
abundances of several elements for different ISM components found in the disk 
and in the halo. The value we find for silicon is the one they quote for 
the warm 
diffuse gas in the halo. In this component, most of the silicon is in the double
ionized form (Si\ioniii/Si\ionii$=2$ -- 14). The ionization potentials of
Fe\ionii, Mn\ionii\ and Mg\ionii\ being lower than that of Si\ionii, it is 
expected
that in cloud D, Fe\ioniii, Mn\ioniii\ and Mg\ioniii\ are at least two times
more abundant than Fe\ionii, Mn\ionii\ and Mg\ionii. The gas phase abundances
we derive for iron, manganese and magnesium are thus, as for silicon, the same
as those quoted for the halo gas in the Fig. 6 of Savage \& Sembach (1996a).

In cloud C, the silicon and the iron gas phase abundances are at least 0.12 and
0.13 dex higher respectively than the values of the depleted low velocity gas
in the disk (Jenkins \& Wallerstein 1996). The silicon gas phase abundance
corresponds to that of the warm gas in the disk in Fig. 6 of Savage \&
Sembach (1996a) and is within the error bars of that of the halo gas. In cloud
C, silicon is mostly in its singly ionized form. We thus assume the same
is true for iron, manganese and magnesium, for which as silicon, the gas phase 
abundances thus also correspond to those of the warm gas in the disk and are
within the error bars of those of the halo gas. 

For the other elements, gas phase abundances were derived for the total
sight-line only.

The abundance of nitrogen, [N/H]$=-0.25^{+0.24}_{-0.15}$, is in good agreement
with previous values derived from Copernicus observations toward many stars~:
$-0.26^{+0.24}_{-0.34}$ (Ferlet, 1981), and $-0.29^{+0.15}_{-0.10}$ 
(York et al., 1983), as well as with the recent determination by Meyer et al.
(1998a) with the GHRS at high resolution~: $-0.17^{+0.02}_{-0.03}$.

For oxygen, our lower limit [O/H]$\geq-0.52$ is compatible with Copernicus 
results of [O/H]$=-0.39^{+0.24}_{-0.10}$ (York et al. 1983, Keenan et al. 1985) 
and very close to the Meyer et al. (1998b) accurate determination of 
$-0.43\pm0.02$, which is only $0.09$ higher.
If Meyer's value is applicable to the \bcma\ sight-line
this implies that its O\ioni\ column density should not exceed 
$7.3\ 10^{14}$~\cmd.

For carbon, the lower limit derived from the strongly saturated C\ionii\
line does not give any interesting constraint.

The aluminium abundance in the sight-line of \bcma\ is known with a rather good
precision. Aluminium  is found to be significantly depleted, but less 
than toward $\alpha$ Vir (York \& 
Kinahan 1979) or toward $\lambda$ Sco (York 1983). Jenkins \& Wallerstein (1996)
show in their Table 5 the aluminium abundance in the gas phase for the depleted
low velocity gas in the disk for \dens=3~\cmt\ and \dens=0.1~\cmt. 
These values are respectively $\sim1.7$ dex and $\sim0.7$ dex lower 
than that derived from our data. Conversely the \bcma\ value agrees 
with those found toward 
halo stars like HD 18100 (Savage \& Sembach, 1996b) or HD 72089, HD 22586, 
HD 49798 and HD 120086 (Jenkins \& Wallerstein, 1996).

Therefore it appears that both clouds have gas phase abundances 
corresponding to warm
gas. Component C abundances are compatible with those of warm clouds in the 
disk, and with those of halo clouds if the amounts of 
Fe\ioniii, 
Mg\ioniii\ and Mn\ioniii\ are significant. 
Component D shows a depletion pattern similar to that of the halo
clouds.
\section{Similarity to other ISM components and nature of the gas}
Table 6 summarizes the characteristics of components C and D derived 
from the observations.
\begin{table}
\caption[]{C and D cloud characteristics derived from the 
observations.}
%\begin{center}
\begin{tabular}{lll}
\hline
\noalign{\smallskip}
cloud & C & D\\
\noalign{\smallskip}
\hline
\noalign{\smallskip}
$V_{\rm helio}$ (\kms) & 14.5 & 25\\
\noalign{\smallskip}
$V_{\rm LSR}$ (\kms) & $-5.0 $ & 5.5\\
\noalign{\smallskip}
{\it T} (K) & $\leq9500$ & $\leq13\,500$\\
\noalign{\smallskip}
$V_{\rm turb}$ (\kms) & 3.2 -- 4.0& 4.0 -- 4.8\\
\noalign{\smallskip}
$N({\rm H}_{\rm tot}$) (\cmd) & 3.7 -- 8.0\ 10$^{18}$ & 1.0 -- 1.6\ 10$^{19}$\\
\noalign{\smallskip}
{\it N}(H\ioni)/{\it N}(\htot) & $\leq 0.60$ & $\leq 0.07$\\
\noalign{\smallskip}
\dens\ (\cmt) $^{1}$ & 0.18 -- 0.54 & 0.38 -- 1.82\\
\noalign{\smallskip}
\hline
%\multicolumn {3}{l}{$^{1}$ {\it \small from Mg\ioni/Mg\ionii, assuming 
%ionization equilibrium}}\\
%&  {\it \small and {\it T}=7000~K}&\\
\end{tabular}\\
$^{1}$ {\small from Mg\ioni/Mg\ionii, assuming ionization equilibrium and {\it
T}=7000~K}
%\end{center}
\end{table}

Warm diffuse ionized gas is a major component of the interstellar medium 
in our Galaxy. It is detected mostly through pulsar dispersion measures 
and H$\alpha$ emission, especially at high Galactic latitudes. 
It is warm ($T\sim 10\,000$~K) and like our components it 
presents line ratios which differ from that of classical H\ionii\ regions
(e.g. Reynolds 1991).  The warm ionized gas is often seen to be 
associated with neutral gas 
as in the line of sight of HD\,93521 where Spitzer \& Fitzpatrick (1993)
observed clouds where neutral and ionized gas seem to be mixed in
partially ionized clouds, as in component C.
Reynolds et al. (1995) have compared H$\alpha$ and 21 cm emission 
in a
$10^{\circ}\,\times\,12^{\circ}$ region of the sky away from the galactic
plane and found that 30\% of the H$\alpha$ emitting gas is spatially and
kinematically associated with 21~cm emitting gas but show that in this case 
H$^0$ and H$^+$ gas occupy close but separate regions. Here again the 
temperature of these "H$\alpha$
emitting H\ioni\ clouds" is about 8000~K, with  electron densities 
around 0.2~\cmt\ and hydrogen ionization ratios 
({\it N}(H\ioni)/{\it N}(\htot)) ranging
from 0.5 to 0.9. All these characteristics are comparable to 
those derived here, especially for component C.

The ionization
mechanism of the diffuse ionized gas is still unclear (e.g. Reynolds et al.
1995). Many authors (Domg\"{o}rgen \& Mathis 1994, Miller \& Cox 1993, 
Reynolds \&
Ogden 1979, Mathis 1986) invoke photoionization by O stars but often note that
the fact that the UV photons can reach the diffuse gas at high latitude many
parsecs away without being absorbed remains a mistery. Ogden \& Reynolds
(1985) use a model of a shock propagating through the ambient medium to
explain the ionization of the long filament they have observed. Spitzer \&
Fitzpatrick (1993) mention penetrating radiation (energetic particles, X-rays) 
or in situ sources (dark matter decay, Sciama 1990) to explain the ionization 
of the H$^+$ and H$^0$ mixed clouds. 
Whether components C and D observed
in the line of sight of \bcma\ are of comparable nature 
as the widespread warm ionized gas remains an open question.

Our components can also be compared to other interstellar constituents
by their elemental abundances.
Component D shows abundances similar to that of 
the high velocity halo clouds and to that of supernova remnants observed toward
HD 72089 (Jenkins \& Wallerstein 1996) which are known to have been shocked. 
As noted by Savage \& Sembach (1996a), shocks are
the most important process in the destruction of dust grains, and the low
depletion in the halo clouds should result from the processes that inject
the clouds into the halo. We suggest that cloud D could 
have been subjected to shocks in the past. This suggestion is supported 
by the ionization ratios which are compatible with collisional
ionization and recombination after cooling 
(as in Trapero et al. 1996). 

For cloud C the situation is somewhat less clear, as it is slightly 
more depleted 
than cloud D and less ionized. However cloud C has abundances similar to 
that of the warm disk and halo gas. It could also have been shocked, 
as it has ionization ratios close to those given by 
collisional ionization at a temperature 
lower than that derived for cloud D. 

The characteristics of cloud C and D are also 
comparable to those of other local interstellar clouds. 
The electron density of the LIC derived by Gry et al. (1995) from the 
Mg\ionii/Mg\ioni\ ratio is $n_e\,=\,.09^{+.23}_{-.07}$~\cmt\ and the 
central value corresponding to the prefered
temperature of $T\sim7000$~K implies a hydrogen ionization fraction close to 
50\%. Wood \& Linsky (1997) have derived a similar range for the electron 
density of the LIC in the line of sight of Capella,
$n_e\,=\,.11^{+.12}_{-.06}$~\cmt, from
the excited state C\ionii\ column density, confirming that hydrogen is 
substantially ionized in the LIC.

A study of elemental abundances in the LIC has been performed through 
the analysis of new GHRS Ech-A data for \ecma\ by 
Gry \& Dupin (1997) and suggests 
that the Local Cloud shows very little depletion, implying that 
most of the matter has been returned to the gas phase
by some dust sputtering process. Another small component in the line of sight 
of \ecma\ has been analysed with the same data and its preliminary study
has been presented by Dupin \& Gry (1997).
As in the case of our component D in the line of sight of \bcma, 
this component is warm ($\sim8000$~K), strongly ionized 
(more than 95\%), and only slightly depleted. It also has ionization 
ratios compatible with those given by collisional equilibrium at 
$T\sim25\,000$~K. 

These results show that this overall region 
of the local interstellar medium is substantially ionized and undepleted.
The characteristics of the components and their similarities with 
diffuse high velocity halo clouds 
lead us to suggest that the Canis Majoris 
tunnel was affected by violent events in the past,
perhaps related to the Local Bubble formation.
\section{Conclusions}
We have studied the nature of the ionized gas that dominates
the \bcma\ sight-line. There appears to be two distinct 
components (named C and D)
with ionization fractions greater than 40\% and 90\% respectively. 
%These
%components are local (between 3 and 150 pc), have $T\le9500$~K and 
%$T\le13\,500$~K respectively, and are diffuse with \dens=0.18 -- 0.54~\cmt\ 
%and \dens=0.38 -- 1.82~\cmt\ for typical warm diffuse clouds temperatures. 
Constraints have been derived for 
the depletion which are similar to those for the warm diffuse gas in both
the disk and halo
for component C, and to those for the high velocity halo clouds for component D.
Silicon seems to be undepleted in cloud D and only slightly depleted in 
cloud C, which is also the case for the other elements studied.

The ionization ratios observed are roughly compatible
with those given by theoretical calculations of collisional ionization 
equilibrium at temperatures close to 20\,000~K.
As the kinetic temperatures of the clouds are actually a factor of two lower,
we conclude that if the clouds are collisionally ionized, they are out of 
equilibrium. Nevertheless photoionization 
by \bcma\ and \ecma\ has not been ruled out and 
the data we present should be analysed with more complete photoionization 
models taking into account the strong EUV excess of these stars.

The processes which have led to this physical state could be the same 
as those which have given rise to 
the high velocity clouds detected in the halo. New obervations of the 
interstellar medium in the sight-line to \ecma\ lead to similar conclusions 
for the LIC and 
another component detected toward this star. A shock which travelled in the 
past through these clouds could have ionized them and disrupted the 
dust grains giving rise to the ionization and low depletion values found ; the 
clouds would now be recombining after cooling. Thus, the main 
components detected towards
\bcma\ could result from the blast-wave that created the 
so-called Local Bubble.
%
%________________________________________ Do not leave a blank line here!
%
\begin{acknowledgements}
We thank Alfred Vidal-Madjar for making the data available to us and 
Dan Welty for providing his profile fitting software. 
We are grateful to Alan Harris and to our referee K. de Boer for suggestions 
which led to substantial improvements on the manuscript.
\\
\end{acknowledgements}

\end{document}